\documentclass[aps,prl,floatfix,twocolumn,reprint,amsmath,amssymb,superscriptaddress,UTF8]{revtex4-2}
\usepackage{amsfonts}
\usepackage{mathrsfs}
\usepackage{amsmath}
\usepackage{color}
\usepackage{natbib}
\usepackage{graphicx}
\usepackage{bm}
\usepackage{amssymb}
\usepackage{xspace}
\usepackage{epstopdf}
\usepackage{dcolumn}
\usepackage{longtable}
\usepackage{array}
\usepackage{makecell}
\usepackage{tabulary}
\usepackage{multirow}
\newcolumntype{C}[1]{>{\centering\arraybackslash}p{#1}}
\newcolumntype{L}[1]{>{\flushleft\arraybackslash}p{#1}}

\usepackage{multirow}
\usepackage{epsfig}
\usepackage[normalem]{ulem}
\usepackage{amsmath}
\usepackage{braket}
\usepackage{bbold}
\usepackage{natbib}

\usepackage[colorlinks=true, letterpaper=true, pdfstartview=FitV, linkcolor=blue, citecolor=blue, urlcolor=blue]{hyperref}

\makeatletter

\newcommand{\Rmnum}[1]{\expandafter\@slowromancap\romannumeral #1@}
\makeatother

\begin{document}

\title{Switchable Giant Spin Injection Current in Janus Altermagnet Fe$_2$SSeO}

\author{Fanxian Pei}
\affiliation{Key Lab of advanced optoelectronic quantum architecture and measurement (MOE), Beijing Key Lab of Nanophotonics $\&$ Ultrafine Optoelectronic Systems, and School of Physics, Beijing Institute of Technology, Beijing 100081, China}
\affiliation{International Center for Quantum Materials, Beijing Institute of Technology, Zhuhai, 519000, China}

\author{Run-Wu Zhang}
\email{zhangrunwu@bit.edu.cn}
\affiliation{Key Lab of advanced optoelectronic quantum architecture and measurement (MOE), Beijing Key Lab of Nanophotonics $\&$ Ultrafine Optoelectronic Systems, and School of Physics, Beijing Institute of Technology, Beijing 100081, China}
\affiliation{International Center for Quantum Materials, Beijing Institute of Technology, Zhuhai, 519000, China}

\author{Lei Li}
\email{lilei1993@imu.edu.cn}
\affiliation{Research Center for Quantum Physics and Technologies, Inner Mongolia University, Hohhot 010021, China}
\affiliation{Inner Mongolia Key Laboratory of Microscale Physics and Atomic Manufacturing, Inner Mongolia University, Hohhot 010021, China}
\affiliation{School of Physical Science and Technology, Inner Mongolia University, Hohhot 010021, China}

\author{Dan Li}
\affiliation{School of Physical Science and Technology, Inner Mongolia University, Hohhot 010021, China}

\author{Yugui Yao}
\affiliation{Key Lab of advanced optoelectronic quantum architecture and measurement (MOE), Beijing Key Lab of Nanophotonics $\&$ Ultrafine Optoelectronic Systems, and School of Physics, Beijing Institute of Technology, Beijing 100081, China}
\affiliation{International Center for Quantum Materials, Beijing Institute of Technology, Zhuhai, 519000, China}

\date{\today}
\begin{abstract}
Generating and controlling spin current in miniaturized magnetic quantum devices remains a central objective of spintronics, due to its potential to enable future energy-efficient information technologies. Among the existing magnetic phases, altermagnetism have recently emerged as a highly promising platform for spin current generation and control, going beyond ferromagnetism and antiferromagnetism. Here, we propose a symmetry-allowed spin photovoltaic effect in two-dimensional (2D) altermagnetic semiconductors that enables predictable control of giant spin injection currents. Distinct from parity-time-- ($\mathcal{PT}$)--antiferromagnets, Janus altermagnetic semiconductors generate not only shift current but also a unique injection current with spin momentum locked in a specific direction under linearly polarized light---a mechanism absent in $\mathcal{PT}$--antiferromagnets. Through symmetry analysis and first-principles calculations, we identify Janus Fe$_2$SSeO as a promising candidate. Specifically, the monolayer Fe$_2$SSeO exhibits a polarization-dependent injection conductivity reaching $\sim$1,200~$\mu$A/V$^{2}\!\cdot\!\hbar/2e$, and the giant spin injection current can be effectively switched by rotating the magnetization direction and engineering strains. These findings underscore the potential of 2D altermagnets in spin photovoltaics and open avenues for innovative quantum devices.
\end{abstract}
\maketitle

\textit{\textcolor{blue}{Introduction.}}--
The development of next-generation spintronics~\cite{vzutic2004spintronics, bader2010spintronics} fundamentally depends on resolving a core challenge~\cite{awschalom2007challenges,sinova2015spin,baltz2018antiferromagnetic}: identifying capable materials of efficiently generating and controlling spin currents to enable energy-efficient spin-based technologies.
Over the past decade, many spin-dependent transport phenomena have advanced the field, drawing widespread interest from both fundamental and applied perspectives~\cite{ladd2010quantum,umesh2019survey}. 
Nevertheless, the ultimate goal of achieving highly controllable spin currents remains elusive. 
Current approaches to spin current generation utilize various physical mechanisms, such as the spin Hall effect~\cite{dyakonov1971current,murakami2003dissipationless,sinova2004universal}
 in metals or the quantum spin Hall effect~\cite{kane2005quantum,bernevig2006quantum,li2025planar}
 in topological insulators. However, these methods suffer from two major limitations: the first concerns the narrow range of candidate materials, which are typically restricted to nonmagnetic heavy metals with strong spin-orbit coupling, ferromagnets that produce stray fields, or parity-time-- ($\mathcal{PT}$)-symmetric antiferromagnets~\cite{cheng2014spin,zhang2014spin,wang2014antiferromagnonic,frangou2016enhanced,rezende2016diffusive,jungwirth2016antiferromagnetic,baltz2018antiferromagnetic,lebrun2018tunable,manchon2019current,hou2019spin}
 with spin polarizations that are difficult to harness; the second involves their general reliance on electrode contacts and their limited response times, typically on the nanosecond scale or slower. The former severely constrains material selection, excluding systems based on light elements or unconventional spin-splitting antiferromagnets. More importantly, the latter---namely the inherent need for electrical contacts and intrinsically restricted switching speeds---currently precludes the generation of spin currents in a contact-free, ultrafast manner.

Recent advances~\cite{hayami2019momentum,vsmejkal2022emerging,zhu2023topological,bai2024altermagnetism,lee2024broken,zhou2024crystal,krempasky2024altermagnetic,reimers2024direct,zhang2024predictable,he2024quasi,zhang2025tunable} in altermagnetic materials
 provide a promising pathway to overcome these challenges. Uniquely positioned at the intersection of desirable electronic properties, intrinsic lattice anisotropy, and efficient coupling with optical fields, altermagnets merge the advantages of ferromagnets and $\mathcal{PT}$--antiferromagnets while circumventing their key limitations, including dependence on heavy elements, unwanted stray fields, and inefficient spin utilization. More significantly, through the nonlinear photogalvanic effect~\cite{ganichev2002spin,bhat2005pure,zhou2007deduction,young2013prediction,xu2021pure,fei2021pt,xiao2021spin}, which is governed by distinctive time-reversal symmetry ($\mathcal{T}$) and $\mathcal{PT}$ symmetry~\cite{wang2019ferroicity,ahn2020low,watanabe2021chiral,nagaosa2024nonreciprocal} properties that give rise to shift and injection currents~\cite{von1981theory,sipe2000second}, altermagnets enable contact-free, ultrafast spin current generation with tailored optical polarization. These attributes establish altermagnets, especially two-dimensional (2D) altermagnets, not only expand the material paradigm for desired materials but also pave the way for optical spintronic devices operating without electrical contacts at unprecedented speeds.

In this work, we propose a distinct spin photogalvanic effect (SPVE) that enables predictable control of giant spin injection currents in 2D Janus altermagnets, simultaneously overcoming the two major limitations of conventional spin current generation schemes. We first elucidate the fundamental differences in spin photocurrent generation mechanisms between $\mathcal{PT}$--antiferromagnets and altermagnets in two dimension. Unlike $\mathcal{PT}$--antiferromagnets, altermagnets support not only shift currents but also unique injection currents exhibiting specific spin-momentum locking under linearly polarized light illumination. We further demonstrate that Janus Fe$_2$SSeO monolayer hosts switchable giant spin injection currents through magnetization rotation and strain engineering. Remarkably, the Fe$_2$SSeO monolayer achieves a polarization-dependent injection conductivity of approximately $1200~\mu$A/V$^{2}\!\cdot\!\hbar/2e$. Our findings not only provide desired material platform for exploring SPVE but also offer new opportunities for developing low-dimensional altermagnetic quantum devices.

\begin{figure}
\includegraphics[width=1.0\columnwidth]{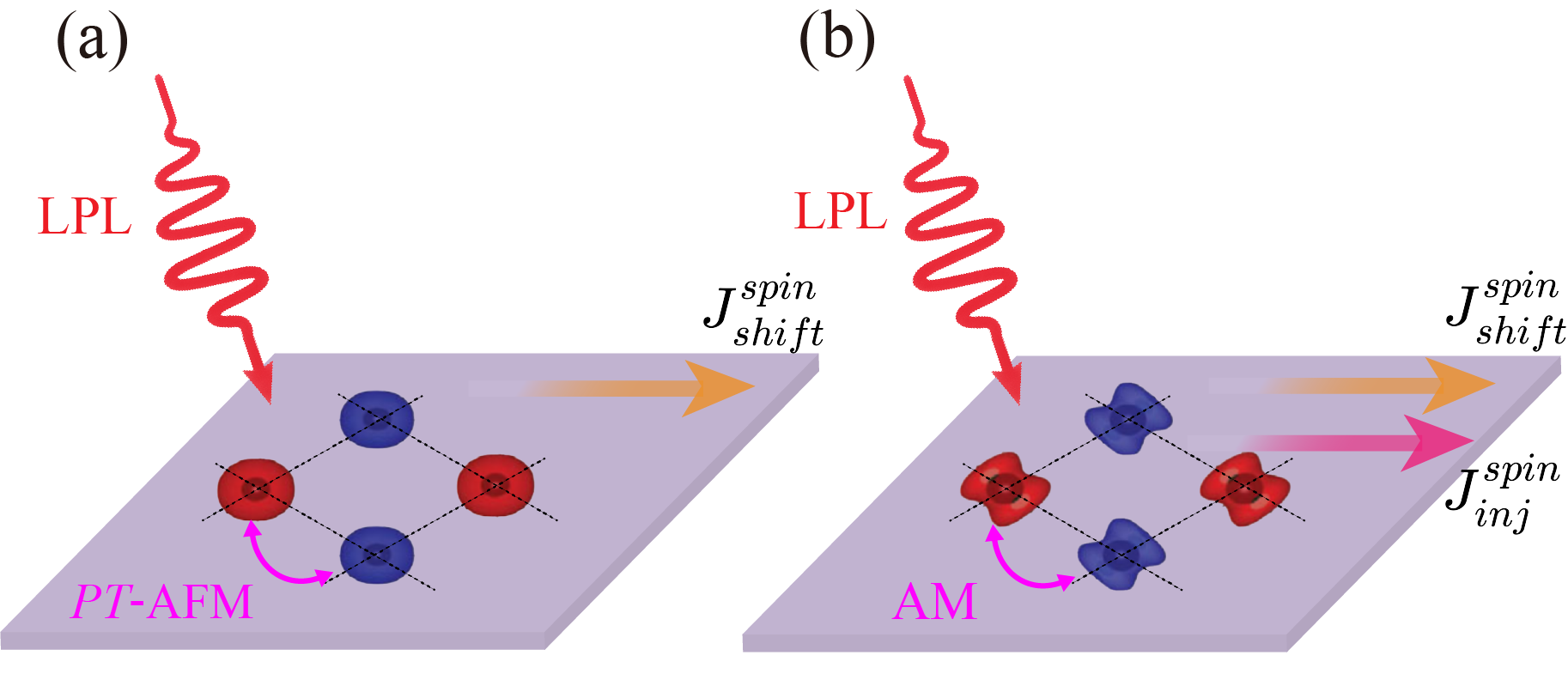}
\caption
{Schematic of spin current generation by linearly polarized light (LPL) in (a) $\mathcal{PT}$-AFM and (b) AM. Unlike $\mathcal{PT}$-AFM, which is restricted to a spin shift current by $\mathcal{PT}$-symmetry, AM breaks $\mathcal{P}$, $\mathcal{T}$, and $\mathcal{PT}$, enabling both spin shift and spin injection currents simultaneously.}
\label{fig1}
\end{figure}

\textit{\textcolor{blue}{Spin injection currents in 2D altermagnets.}}--
The SPVE refers to the generation of a pure spin current under light illumination. 
When a material is irradiated with light of frequency $\omega$, nonlinear optical (NLO) processes can generate a direct-current spin current described by the following expression:
\begin{align}\label{Eq:SPVE}
j_\mathrm{dc}^{c,s^i} & =\sum_{\Omega=\pm\omega}\sigma_{ab}^{c,s^i}(0;\Omega,-\Omega)E_a(\Omega)E_b(-\Omega),
\end{align}
where $E(\omega)$ is the Fourier component of the electric field at angular frequency $\omega$, $\sigma$ denotes the NLO spin conductivity, $c$ indicates the flow direction of the spin current, $a$ and $b$ specify the polarization directions of the electric field, and $s^i$ ($i = x, y, z$) labels the spin polarization direction.
Akin to the bulk photovoltaic effect (BPVE)~\cite{belinicher1980photogalvanic,von1981theory}, which comprises shift and injection current contributions, the SPVE can similarly be decomposed into these two components. The shift current arises from the change in electron position during interband transitions, while the injection current originates from the change in electron velocity. 
Under a point-group symmetry operation $R$, a vector transforms as $r_a \rightarrow r^{\prime}_a = \sum_{b} \mathcal D(R)_{ab} r_b $~\cite{gallego2019automatic}. Accordingly, the transformation~\cite{ahn2020low,holder2020consequences}
 of the fourth-rank spin conductivity tensors under linearly polarized light in Eq. (\ref{Eq:SPVE}) is given by 
\begin{align}\label{Eq:symmetry}
	{\sigma^{\prime}}_{\text{shift,L}}^{c_1,s^{i_1};a_1b_1} &= (-1)^{s_T} \det(R) \sum_{a,b,c,i} \mathcal D(R)_{c_1c} \mathcal D(R)_{i_1i} \notag \\
	&\quad \times \mathcal D(R)_{a_1a} \mathcal D(R)_{b_1b} \sigma_{\text{shift,L}}^{c,s^{i};ab}, \notag \\
	{\sigma^{\prime}}_{\text{inj,L}}^{c_1,s^{i_1};a_1b_1} &= \det(R) \sum_{a,b,c,i} \mathcal D(R)_{c_1c} \mathcal D(R)_{i_1i} \mathcal D(R)_{a_1a}  \notag \\
	&\quad \times \mathcal D(R)_{b_1b} \sigma_{\text{inj,L}}^{c,s^{i};ab}.
\end{align}
Note that for a symmetry operation that includes time reversal, we have $(t,r_a) \rightarrow (t^\prime, r^\prime_a) = ((-1)^{s_T}t, \sum_{b} \mathcal D(R)_{ab} r_b)$. The factor of determinant $\det(R)$ appears in above transformation of spin-related quantities because spin is an axial vector.

As is evident from Eq. (\ref{Eq:symmetry}),  broken spatial inversion symmetry ($\mathcal{P}$) is necessary for the SPVE to occur, while the specific form of the resulting spin current---whether shift or injection mechanism---is determined by the material's magnetic structure. Specifically, $\mathcal{PT}$ symmetry will prohibit the tensor $\sigma_{\text{inj,L}}^{c,s^{i};ab}$ while permitting  $\sigma_{\text{shift,L}}^{c,s^{i};ab}$. The explicit form of these conductivity tensors are provided in Section II of Supplementary Material (SM). In $\mathcal{PT}$--antiferromagnets, the spin-up and spin-down sublattices are related by the combined $\mathcal{PT}$ operation, which allows the generation of a shift spin current under linearly polarized light.
In altermagnets, by contrast, the opposite spin sublattices are connected by mirror-rotation ($\mathcal{O}$) symmetries. This breaks $\mathcal{P}$, $\mathcal{T}$, and $\mathcal{PT}$ symmetries, allowing both shift and injection currents to emerge simultaneously under linear polarization, as depicted in Fig.~1.

Moreover, in real 2D magnetic systems, the Mermin-Wagner theorem~\cite{mermin1966absence,halperin2019hohenberg}
 prohibits spontaneous symmetry breaking at finite temperatures, in contrast to their three-dimensional (3D) counterparts.
This fundamental constraint makes a relativistic analysis essential for understanding the SPVE in 2D altermagnets. Within a relativistic framework, symmetry analysis must incorporate magnetic symmetry operations that act concurrently on both spin and real-space coordinates. These operations define the constraints on the allowable components of the conductivity tensor $\sigma_{\text{shift/inj}}^{c,s^i;ab}$. To build a symmetry-guided theory for the SPVE in 2D altermagnets, we screened the 125 magnetic layer point groups~\cite{wang2025type}, identifying 30 potential candidate targets. We then systematically classified the groups capable of supporting the SPVE to pinpoint candidate materials possessing the desired physical properties, as summarized in Table S1 of the SM.

More importantly, as the order parameter for altermagnets, the N\'eel vector provides a robust and non-volatile means of controlling magnetic symmetry. Consequently, the SPVE response remains strongly dependent on the orientation of the N\'eel vector and varies periodically as the vector is rotated. Guided by our theoretical framework, we identified Fe$_2$SSeO as a representative altermagnetic system to validate the realization of a tunable giant spin injection current via the SPVE.

 \begin{figure}
 	\includegraphics[width=1.0\columnwidth]{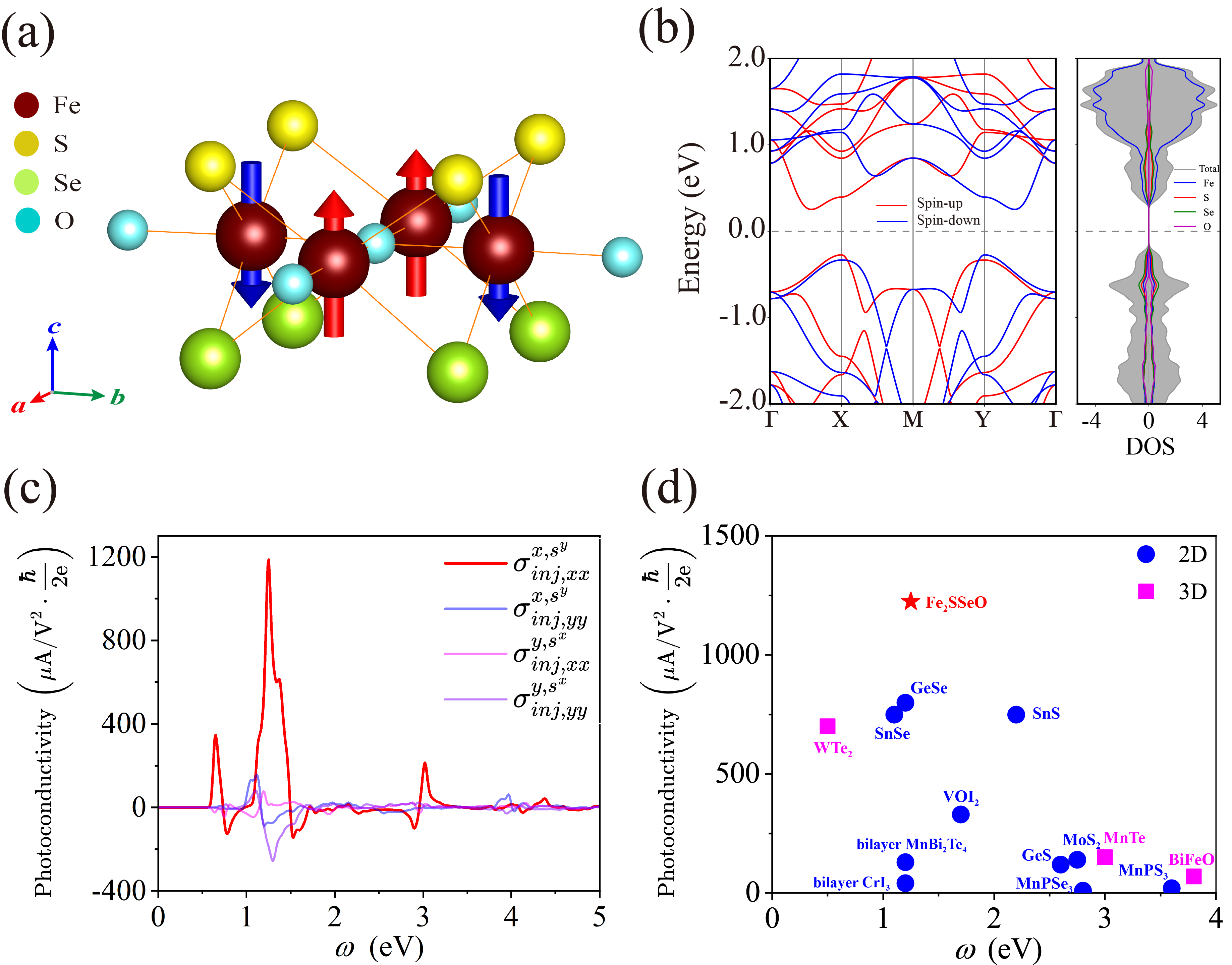}
 	\caption
 	{(a) Crystal structure of Fe$_2$SSeO monolayer  in the N\'eel type AFM configuration, with spin directions indicated by red (up) and blue (down) arrows. (b) Calculated band structure (without SOC) and spin-resolved PDOS for Fe$_2$SSeO monolayer , where red and blue denote spin-up and spin-down components, respectively. Regarding Fe$_2$SSeO monolayer, (c) Shown is the injection spin photoconductivity spectra computed with SOC for the N\'eel vector aligned along the [100] direction.(d) Performance comparison of the maximum spin current conductivity and its photon energy for Fe$_2$SSeO relative to selected 2D and 3D representative materials.}
 	\label{fig2}
 \end{figure}
 
\textit{\textcolor{blue}{Material candidates.}}--
In physics, many promising theoretical concepts have achieved significant progress, yet their practical implementation often hinges on the identification of suitable material candidates.
The recent discovery of room-temperature metallic \textit{d}--wave altermagnetism in layered vanadium oxytellurides Rb$_{1-\delta}$V$_2$Te$_2$O and KV$_2$Se$_2$O~\cite{ablimit2018weak,zhang2025crystal,jiang2025metallic}, combined with the successful experimental synthesis of layered iron oxychalcogenides BaFe$_2$Se$_2$O and NaFe$_2$S$_2$O~\cite{he2011synthesis,lei2012structure,takeiri2016high,song2023crystal}, highlights the promise of this material family. Their pronounced van der Waals character motivates our investigation into their Janus monolayer derivatives.

Crystal structure of Janus Fe$_2$SSeO monolayer is illustrated in Fig.~2(a). The structure consists of a middle layer formed by transition metal Fe and oxygen atoms, sandwiched between a top S layer and a bottom Se layer, forming a characteristic Janus trait. The Janus Fe$_2$SSeO monolayer (space group No.99, $P4mm$) exhibits reduced symmetry due to broken out-of-plane mirror symmetry, while retaining tetrahedral coordination and in-plane $\mathcal{C}_{4}$ rotational symmetry. This material system has well-defined structural origins, and its single crystalline form can be derived from layered parent compounds. Using Fe$_2$SSeO as a representative example, we systematically investigate the SPVE and its modulation mechanisms in 2D Janus altermagnets.

\begin{table*}
\caption{\label{tab:table3}Summary of the symmetry-allowed nonzero components for the linear shift and injection currents in Fe$_2$SSeO, evaluated for light polarization  along the $L \parallel \hat{x}$, $L \parallel \hat{y}$, and $L \parallel \hat{z}$ directions.
}
\label{tab:spin_current}
\begin{ruledtabular}
\renewcommand{\arraystretch}{1.5}
\begin{tabular}{ccccc}
 N\'{e}el vector&Linear shift current&Linear injection current\\ \hline
 $L \parallel \hat{x}$&$\sigma_{xx}^{xs^z}$,$\sigma_{yy}^{xs^z}$
 &$\sigma_{xx}^{xs^y}$,$\sigma_{yy}^{xs^y}$,$\sigma_{xx}^{ys^x}$,$\sigma_{yy}^{ys^x}$ 
  \\
 $L \parallel \hat{y}$&$\sigma_{xx}^{ys^z}$,$\sigma_{yy}^{ys^z}$
 &$\sigma_{xx}^{xs^y}$,$\sigma_{yy}^{xs^y}$,$\sigma_{xx}^{ys^x}$,$\sigma_{yy}^{ys^x}$
 \\
 $L \parallel \hat{z}$&$\sigma_{xx}^{xs^x} = -\sigma_{yy}^{ys^y}$,$\sigma_{yy}^{xs^x} = -\sigma_{xx}^{ys^y}$ 
 &$\sigma_{xx}^{xs^y} = -\sigma_{yy}^{ys^x}$,$\sigma_{yy}^{xs^y} = -\sigma_{xx}^{ys^x}$
 \\
\end{tabular}
\end{ruledtabular}
\end{table*}

The calculated band structure and partial density of states (PDOS) of the Fe\textsubscript{2}SSeO monolayer without considering SOC are shown in Fig.~2(b). Analysis reveals that Fe\textsubscript{2}SSeO is an indirect bandgap semiconductor with a bandgap of 0.526 eV and exhibits significant spin splitting---a hallmark feature of altermagnetism. This behavior breaks the Kramers degeneracy commonly present in conventional $\mathcal{PT}$--antiferromagnets, leading to spin-split energy bands. These distinctive crystal and electronic properties make Fe\textsubscript{2}SSeO an ideal candidate for exploring spin-dependent transport phenomena.

\begin{figure}
	\includegraphics[width=1.0\columnwidth]{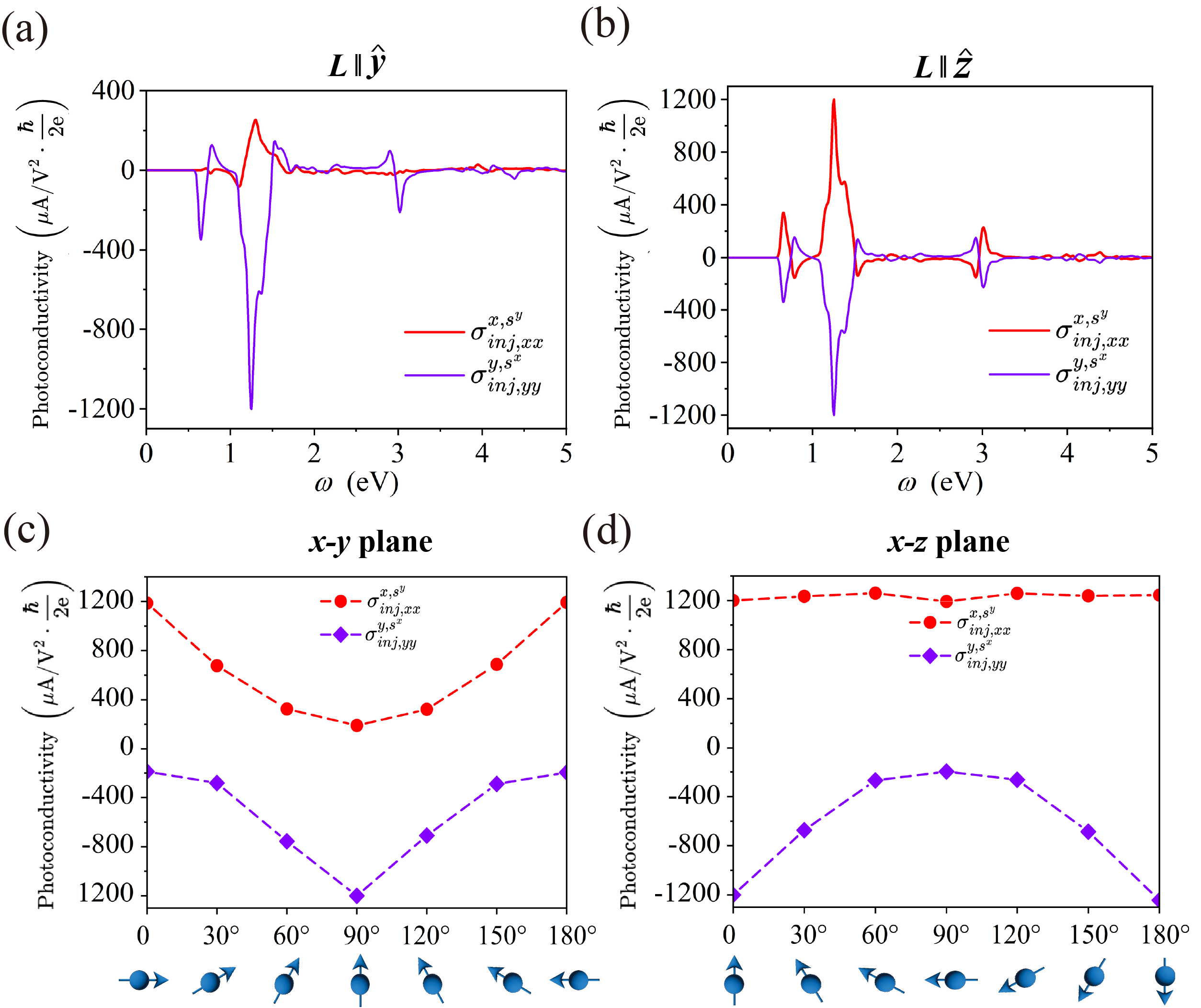}
	\caption
	{(a-b) Injection spin photoconductivity components $\sigma_{xx}^{xs^y}$ and $\sigma_{yy}^{ys^x}$ under linearly polarized light, showcasing responses for polarization vectors along (a) $L \parallel \hat{y}$,(b) $L \parallel \hat{z}$. (c-d) Evolution of the nonlinear optical conductivities $\sigma_{xx}^{xs^y}$ and $\sigma_{yy}^{ys^x}$ in the (c) $x$-$y$ plane and (d) $x$-$z$ plane rotation angle $\alpha$ of the N\'eel vector.
    }
	\label{fig3}
\end{figure}

\textit{\textcolor{blue}{Switchable giant spin injection current.}}-- 
The Janus Fe$_2$SSeO monolayer exhibits SPVE due to broken $\mathcal{P}$ symmetry, enabling nonlinear spin current generation under optical excitation. Magnetic anisotropy calculations confirm soft magnetic behavior and identify an in-plane easy magnetization axis along the [100] direction (see Sec.~IV of the SM), consistent with the magnetic point group ${pm'm2'}$. Applying the material's symmetry operations to Eq. (2) yields independent tensor components for injection and shift spin currents (see Sec.~V of the SM).
Notably, although both spin currents are excited, their nonzero tensor elements are mutually independent and do not overlap. Under linearly polarized light, the independent nonzero components of the injection spin conductivity are $\sigma_{xx}^{xs^{y}}$, $\sigma_{yy}^{xs^{y}}$, $\sigma_{xx}^{ys^{x}}$, and $\sigma_{yy}^{ys^{x}}$. In contrast, the shift spin conductivity is constrained by $\mathcal{T}$ symmetry: the $\mathcal{M}_{y}\mathcal{T}$ operation nullifies the above four components, leaving only $\sigma_{xx}^{xs^{z}}$ and $\sigma_{yy}^{xs^{z}}$ as nonzero. This behavior arises from the coexistence of mirror and $\mathcal{T}$ symmetries in the structure---a feature also observed in BPVE .

The NLO injection spin current conductivity of Fe$_2$SSeO monolayer under linearly polarized light [Fig.~2(c)] shows pronounced spectral features, with the $\sigma_{xx}^{xs^y}$ component exhibiting three distinct peaks in the $\omega = 0\sim5$ eV range. The most prominent peak, located at $\omega = 1.25$ eV, reaches a magnitude of $1224$ $\mu$A/V$^{2}\!\cdot\!\hbar/2e$, significantly exceeding values reported for other widely studied 2D and 3D materials [Fig.~2(d)]~\cite{mu2021pure,xiao2021spin,xu2022abnormal,xue2023valley,liao2024enormous,dong2025crystal,yang2025giant}. Previous analysis indicates that $\sigma_{xx}^{xs^y}$ corresponds to injection current, while the other injection conductivity components $\sigma_{yy}^{xs^y}$, $\sigma_{xx}^{ys^x}$, and $\sigma_{yy}^{ys^x}$ are considerably smaller. As shown in Fig.S3, the shift conductivities $\sigma_{xx}^{xs^z}$ and $\sigma_{yy}^{xs^z}$ under linearly polarized light are also significantly lower than the injection terms, indicating that although both shift and injection spin currents are optically excited in monolayer Fe$_2$SSeO, the injection current dominates.

The SPVE exhibits strong sensitivity to the N\'eel vector orientation, as summarized in the Table~\ref{tab:spin_current} listing the non-zero components of $\sigma$ for different N\'eel vector directions in Fe$_2$SSeO. Additionally, the spin currents generated under different linear polarizations can be interconnected through the unique crystal symmetry $\mathcal{O}$ in altermagnets. When the magnetic axis is aligned along the [010] direction, the the 90\(^\circ\) rotation of the N\'eel vector causes the large spin current originally carried by $\sigma_{xx}^{xs^{y}}$ to transform into the $\sigma_{yy}^{ys^{x}}$ component, accompanied by a sign reversal [Fig.~3(a)]. In contrast, when the magnetic axis is oriented along the [001] direction, the magnetic point group of the system changes to $\mathcal{C}_{4z}$ rotational symmetry, which constrains $\sigma_{xx}^{xs^{y}} = -\sigma_{yy}^{ys^{x}}$, allowing both orientations to support a large spin current [Fig.~3(b)]. Figs.~3(c-d) further show the evolution of $\sigma_{xx}^{xs^{y}}$  and $\sigma_{yy}^{ys^{x}}$ in the \textit{x-y} and \textit{x-z} plane as the N\'eel vector rotates. Owing to the rotational symmetry of the crystal structure, the conductivity varies with a period of $\pi$, satisfying $\sigma(\alpha) = \sigma(\pi - \alpha)$. These results confirm that the spin current is strictly governed by the N\'eel vector orientation.

From the perspective of application, spin currents can be detected using established methods. For instance, in the spin Hall effect~\cite{sinova2015spin}, an open-circuit configuration leads to spin accumulation at the sample edges, which can be measured via magneto-optical techniques such as Kerr rotation or the Faraday effect~\cite{kato2004observation}. The strong dependence of the SPVE on N\'eel order makes monolayer Fe$_2$SSeO a highly promising candidate for designing SPVE-based spintronic devices.

\begin{figure}
	\includegraphics[width=1.0\columnwidth]{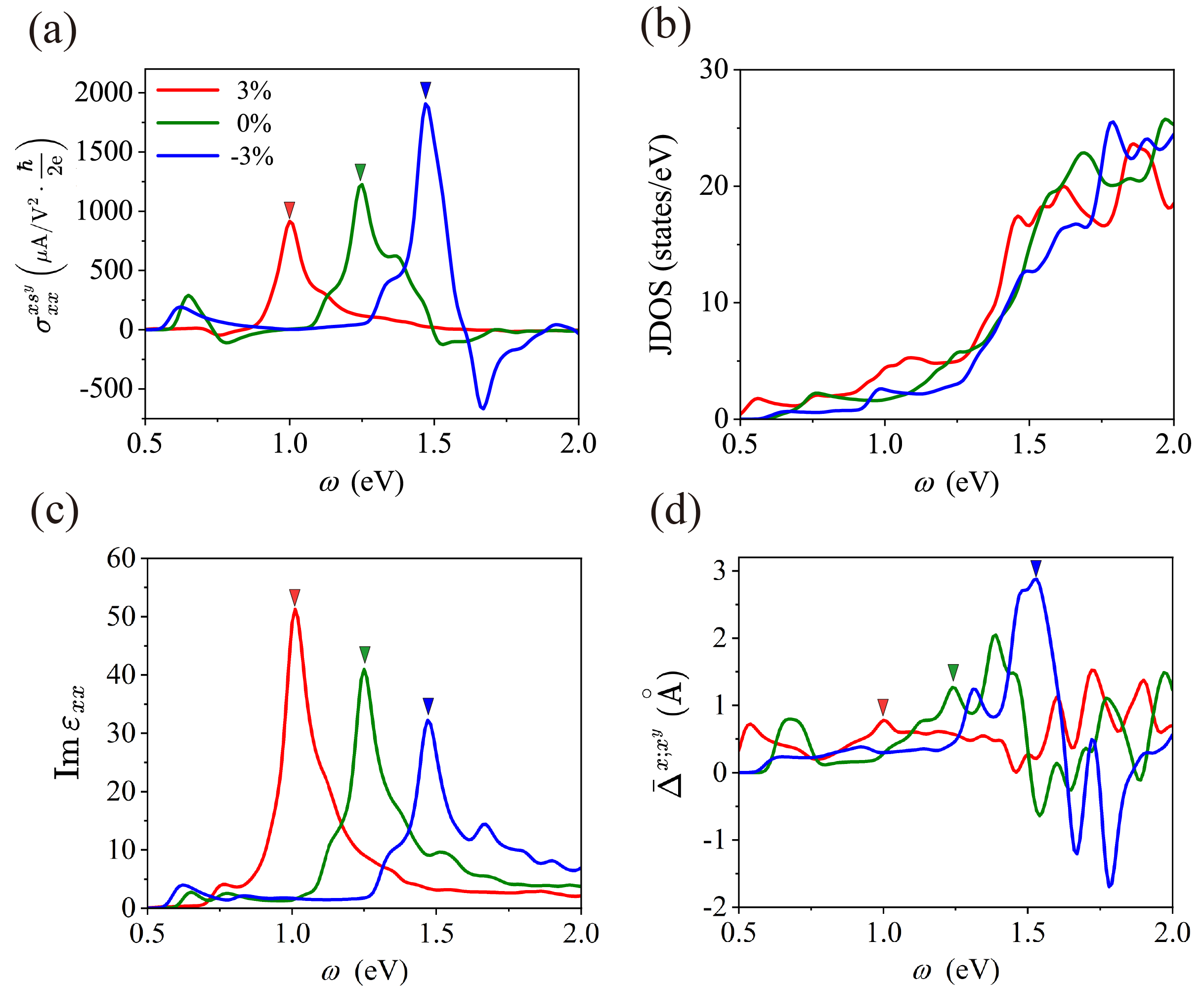}
	\caption
	{(a) Injection spin current conductivity $\sigma_{xx}^{xs^{y}}$, (b) JDOS, (c) the absorptive part of the dielectric function $\operatorname{Im} \varepsilon_{xx}$, and (d) aggregate injection vector $\bar{\Delta}^{x;x^y}$ as a function of the photon energy under different biaxial strain.
	}
	\label{fig3}
\end{figure}

\textit{\textcolor{blue}{Discussion and conclusion.}}--
The development of effective approaches to manipulate nonlinear spin currents is essential for advancing functional material design and device applications. Strain engineering offers a versatile strategy for tuning material properties, applicable to thin films grown on a broad range of substrates ~\cite{nadupalli2019increasing,schankler2021large,dong2023giant,yang2024two}, particularly through substrate-mediated tensile or compressive deformation. 
We further explored the effect of biaxial strain on the SPVE of Fe$_2$SSeO monolayer. As shown in Fig.~4(a), biaxial strain modulates the band structure, leading to distinct responses of the spin current conductivity $\sigma_{xx}^{xs^y}$ under tensile and compressive strains: tensile strain suppresses the conductivity and induces a red shift in the spectral peak, whereas compressive strain enhances the conductivity and leads to a blue shift. Notably, at $-3\%$ compressive strain, the peak conductivity increases from $1224$ $\mu$A/V$^{2}\!\cdot\!\hbar/2e$ ($\omega = 1.25$~eV) to $1905$ $\mu$A/V$^{2}\!\cdot\!\hbar/2e$ ($\omega = 1.47$~eV). To understand the above features, we calculate three additional Brillouin zone integrated quantities: the joint density of states (JDOS) $D_\text{joint}$, the absorptive part of the dielectric function $\varepsilon_\text{ab}$, and the aggregate injection vector $\bar{\Delta}^{c;s^i}$ (detailed definitions are provided in the Sec.II. of the SM). As shown in Figs.~4(c) and 4(d), under strain modulation, both the $\varepsilon_{xx}$ and the $\bar{\Delta}^{x;x^y}$ undergo changes, exhibiting a competitive relationship. However, the variation in the injection vector dominates, leading to a notable enhancement of conductivity under compressive strain and the opposite under tensile strain. This enhancement can be attributed to shortened interatomic distances and a strengthened built-in electric field under compression, which facilitate carrier transport and spin current generation.

In summary, we have established a generalized symmetry-based framework for the SPVE under linearly polarized light in conventional \(\mathcal{PT}\)\textendash antiferromagnets and Janus altermagnets. Our analysis shows that under linearly polarized light excitation, Janus altermagnets support both shift and injection spin photocurrents, whereas only shift currents are allowed in \(\mathcal{PT}\)\textendash antiferromagnets. Using Janus Fe$_2$SSeO monolayer as a representative example, we demonstrated a giant polarization-dependent injection current, with a conductivity of \(1224\ \mu\text{A}/\text{V}^2 \cdot \hbar/2e\) in Fe$_2$SSeO---one to two orders of magnitude larger than typical 2D materials. The SPVE response is highly sensitive to the N\'eel vector orientation, showing \(\pi\)-periodic modulation upon its rotation, and can be further tuned via biaxial strain. The observed giant and switchable spin photocurrent underscores the potential of Janus altermagnets for applications in nonlinear optoelectronics and optical spintronics.


\bibliography{ref}



\end{document}